\documentstyle[12pt,epsfig]{article}

\def\be{\begin{equation}}
\def\ee{\end{equation}}
\def\ba{\begin{eqnarray}}
\def\ea{\end{eqnarray}}
\def\ga{\mathrel{\raise.3ex\hbox{$>$\kern-.75em\lower1ex\hbox{$\sim$}}}}
\def\la{\mathrel{\raise.3ex\hbox{$<$\kern-.75em\lower1ex\hbox{$\sim$}}}}

\newcommand{\bi}[1]{\bibitem{#1}}
\newcommand{\fr}[2]{\frac{#1}{#2}}

\begin{document}

\baselineskip=16pt
\begin{titlepage}

\rightline{UVIC--TH--07/12}
\rightline{UMN-TH-2618/07}
\rightline{FTPI-MINN-07/28}
\rightline{September 2007}
\begin{center}

\vspace{0.5cm}

\large {\bf Environmental Dependence of Masses and Coupling Constants}
\vspace*{5mm}
\normalsize

{\bf Keith A. Olive$^1$} and {\bf Maxim Pospelov$^{2,3}$}

\smallskip
\medskip

${}^1${\it William I.\ Fine Theoretical Physics Institute,\\
University of Minnesota, Minneapolis, MN~55455, USA}

${}^2${\it Department of Physics and Astronomy,  University of Victoria,
Victoria, BC, V8P 1A1 Canada}

${}^3${\it Perimeter Institute for Theoretical Physics, Waterloo,
Ontario N2J 2W9, Canada}

\smallskip
\end{center}
\vskip0.6in

\centerline{\large\bf Abstract}
We construct a class of scalar field models coupled to matter that lead to the dependence of 
masses and coupling constants on the ambient matter density. Such models predict a deviation
of couplings measured on the Earth from values determined in low-density 
astrophysical environments, but do not necessarily require the 
evolution of coupling constants with the redshift in the 
recent cosmological past.  Additional 
 laboratory and astrophysical tests of $\Delta \alpha$ and  $\Delta(m_p/m_e)$ as functions
 of the ambient matter density are warranted. 

\vspace*{2mm}

\end{titlepage}

\section{Introduction}
\setcounter{equation}{0}

Perhaps the most astonishing fundamental observation of the last decade 
was the discovery of dark energy.
So far, all cosmological data are consistent with the simplest 
possibility: dark energy is just a new fundamental constant of nature, which does not evolve 
over cosmological redshifts. On the other hand, it is intriguing to think about 
alternative explanations associated with this profound change in infrared physics.
The most straightforward way of implementing such a change is the  
introduction of a new ultra-light scalar degree of freedom associated with
quintessence \cite{Ratra}. The coupling of this scalar field to matter may be the source of new 
cosmological phenomena such as an apparent ``breakdown" of Lorentz invariance connected to the 
CMB frame, the existence of a ``fifth force" mediated by scalar exchange, 
or a change of couplings and masses with time. Thus, the search for these exotic effects
acquires a new actuality in the ``dark energy" age. 

A potential hint on the difference between the laboratory values of the fine-structure 
constant $\alpha$ \cite{seattle,Gabrielse}
and the one derived from 
quasar absorption spectra at high redshifts \cite{Webb}, $\Delta \alpha / \alpha \sim -0.6 \times 10^{-5}$
 ($z\sim 0.5-3$), has triggered a series of new phenomenological and theoretical 
 studies of ``changing couplings" \cite{massgrave}-\cite{LOP}. 
Subsequent studies that employed the same ``many multiplet method" 
have shown $\Delta \alpha$ consistent with zero with the same accuracy \cite{Petitjean}
(however, a critical discussion of this result is found in \cite{anti-Petitjean}). 
In a recent development, the comparison of H$_2$ 
spectra obtained in the laboratory and at high redshifts, yielded the first hint of a 
possible change of $m_p/m_e$ at the three sigma level \cite{Petitjean2}, whereas 
the analysis of the inversion spectrum of ammonium at $z= 0.68$ led only to an upper 
limit on $\Delta(m_p/m_e)$ \cite{Flambaum:2007fa}. Null results for 
time evolution of couplings are reinforced by the chemical composition of Oklo rocks \cite{Oklo}
and  meteoritic abundances of rhenium, that provide stringent constraints 
and go back to $z\sim 0.2- 0.4$ \cite{OPQ,fuj}. 
Finally, a look into the much deeper
past, to the time of the Big Bang Nucleosynthesis (BBN) \cite{BBNalpha,CO,OS}, leads to the conclusion that the 
couplings could be different from present day values by no more than a few percent at that time.

Despite the controversial status of the non-zero claim for
$\Delta \alpha/\alpha$ there have been a 
significant number of attempts \cite{massgrave}-\cite{LOP}, \cite{Bek}  - \cite{Uzan},  
to build simple models that could account for possible effects of order $O(10^{-5})$. 
Theoretical models of the variation of other couplings were discussed in Refs. \cite{other,CO}. 
The simplest Lagrangian 
that enables the variation of $\alpha$ was written down by 
Bekenstein \cite{Bek}, and involves the coupling of an 
ultra-light scalar field to the kinetic term in the electromagnetic Lagrangian, 
$\zeta_F\fr{\varphi}{M_*}F_{\mu\nu}F^{\mu \nu }$, 
where $M_*$ is a large energy scale, comparable to the Planck scale and $\zeta_F$ is a constant. 
An attractive consequence of this interaction is the absence of the evolution of the coupling in the 
radiation dominated epoch after the annihilation of electron-positron pairs \cite{Bek}. 
The drift of $\alpha$ in time comes about later during the matter dominated
epoch due to a nonzero, $O(10^{-4}-10^{-3})$, 
electromagnetic contribution 
to the nucleon mass, which supplies a coupling of $\varphi$ to the baryon energy density. The 
 time evolution of 
$\alpha$, including the sign, is calculable in terms of one free parameter, $(\zeta_F/M_*)^2$, and 
the result is a linear dependence of $\alpha$ on the logarithm of the redshift in the matter dominated  
epoch. 
Unfortunately,
between redshifts $z\sim 1$ and $z=0$, $\Delta \alpha/\alpha$ is predicted 
to be far too small in the minimal model of 
Bekenstein; several orders of magnitude below the modern observational capabilities, 
once the fifth force constraints on $\zeta_F/M_*$ are taken into account. In attempt to 
resuscitate this model, it has been suggested that a coupling to dark matter \cite{Damour,OP,SBM}
and/or a self-interaction potential \cite{DZ,Chiba}
of the scalar field can drive its evolution in a much more efficient way than the baryon energy density. 
Inescapably, these models have more 
parameters ({\em i.e.} a nearly {\em arbitrary} potential $V(\varphi)$),
which results in a loss of predictivity. Thus almost any redshift dependence for $\alpha(z)$, including oscillatory behavior,
is possible, and no concrete predictions can be obtained from these theoretical models. If the experimental limits on
$\Delta \alpha$ are enforced, it has proven to be difficult to remain consistent with 
a substantial change in the fine structure constant at $z\sim 1$ and
satisfy Oklo and meteoritic constraints. Still, fairly large classes of models (see {\em e.g.}
Refs. \cite{Wett}, \cite{Fujii}- \cite{LOP}) are known to pass these requirements. 
On the downside, if $V(\varphi)$ supplies the driving force for the scalar field, there 
are no convincing  arguments why $\varphi$ would not evolve deep into the 
radiation domination epoch resulting in vastly different couplings during the BBN.
In addition to these specific difficulties, 
there is a generic naturalness problem that all such models suffer from:
it is difficult to argue that the mass of $\varphi$, as well as other terms in $V(\varphi)$, are protected 
to the scale of $10^{-33}$ eV to allow for the requisite late 
evolution, and at the same time allow the $\varphi$-field to have significant couplings to 
matter fields. Any conceivable loop effect or nonperturbative QCD 
vacuum condensate would tend to induce
mass parameters about twenty orders of magnitude in excess of $10^{-33}$ eV (See {\em e.g.} \cite{Dine}). 

In all models of changing couplings discussed so far in the literature, 
the temporal change of $\alpha$ dominates 
over the possible spatial variation of $\alpha$. Although some suggestions 
were made that the spatial variation may dominate over temporal variations \cite{BarOt},
no explicit models where found to date. 
In Bekenstein-type models, the spatial variation of couplings is caused by 
matter inhomogeneities and thus
follows the profile of the gravitational field. If the coupling of the scalar field to matter is not
stronger  than the matter-gravity coupling, one expects the change in 
$\varphi$ to be less than the variation of the metric.  For example, the difference between 
coupling constants on the surface of the 
Earth an in orbit is not going to exceed $10^{-10}$ and in practice will be  much smaller once 
fifth force constraints are imposed. 
In this paper, we show that such a conclusion is not generic, and there is a whole 
class of models where spatial variations can be more pronounced than the cosmological variations 
in the recent past, opening new possibilities for searching for $\Delta \alpha$ and $\Delta m$
as functions of the matter density.

A key to this proposal is to choose the couplings of a scalar field to matter to be much stronger 
than gravitational. At first glance this would appear to only worsen the problem of a fifth force 
induced through scalar exchange. However, this might not happen if the matter density itself 
leads to the effective suppression of the {\em linear} scalar field coupling to matter \cite{DN,DP},
and/or the range of the scalar-mediated force becomes shorter than the one needed for conventional 
fifth-force experiments \cite{cham1,cham2,MS,cham3}. Models that escape prohibitive 
fifth-force constraints may predict 
spatial variation of $\alpha$ and $m_p/m_e$ that exceed recent temporal variations. 
As we will show in the remainder of this paper, such constructions can be achieved 
if the dynamics of the scalar field in low-density environments is determined by its self-potential, 
while in regions of large overdensities, the dynamics of $\varphi$ is set by its coupling to matter. 
In a large subclass of such models where the matter-$\varphi$ coupling is much stronger than 
gravitational, the temporal evolution happens on the time scales that are much shorter than cosmological.
In these models, the global temporal evolution of 
the scalar field could be finished a
long time ago, and on average the ``cosmological" values of masses and coupling constants 
remain constant in space and time. We denote the values of 
couplings and masses in rarefied low density regions as $\{\alpha_r,~m_r\}$. 
The value of the coupling constants on Earth or in any high-density environment 
can be different and are labeled as $\{\alpha_d,~m_d\}$.
In the next sections we discuss the essential features of the model, determine observational constraints 
on its parameters and argue that independent tests of $\alpha_d-\alpha_r$ between different points 
in space are indeed warranted. 

\section{Scalar field models of $\alpha(\rho)$ and $m(\rho)$}
The starting point for our analysis is the matter-gravity-scalar field action,
\begin{eqnarray}
S_{\phi} = \int d^4x \sqrt{-g} \Biggl\{- \frac{M_{\rm Pl}^2}{2} R + \frac{{M_*}^2}{2}
\partial^{\mu} \phi
\partial_{\mu} \phi  - V(\phi) \nonumber \\
- \sum_i \frac{B_{Fi}(\phi)}{4}F^{(i)}_{\mu\nu}F^{(i)\mu\nu} + \sum_j
[\bar\psi_j iD\!\!\!\!/ \psi_j -
B_j(\phi)m_j\bar\psi_j\psi_j]\Biggr\}, \label{Lagrangian}
\end{eqnarray}
which can be viewed as a generalization of a scalar-tensor theory of gravity. 
In this expression,  $M_{\rm Pl}=(8\pi G_N)^{-1/2}$ is the reduced Planck mass, $\phi$ is a
dimensionless scalar field with $M_*$ being the analogue of the Planck mass in the 
scalar sector. The functions $B_{Fi}(\phi)$ give the 
$\phi$-dependence to the gauge couplings in Standard Model (SM), and 
the sum is extended over all SM gauge groups. $\psi_j$ represents Standard Model
fermions that are coupled to $\phi$ via the functions $B_j(\phi)$.
After performing a $\phi$-dependent rescaling of the matter fields, one is allowed to 
remove the $\phi$-dependence of the kinetic
terms for the SM fermions $\psi_i$ and keep only couplings to the mass terms. 
If needed, the interaction (\ref{Lagrangian}) can be generalized further to include Higgs bosons, 
cold dark matter particles, etc. 

Among couplings to the SM model fields, the couplings to quarks, gluons, photons and electrons 
are the most important. At lower energies, we can abandon the quark-gluon description in 
favor of an effective coupling to nucleons 
and reduce (\ref{Lagrangian}) to a more tractable form,
\begin{eqnarray}
S_{\phi} = \int d^4x \sqrt{-g} \Biggl\{- \frac{M_{\rm Pl}^2}{2} R + \frac{{M_*}^2}{2}
\partial^{\mu} \phi
\partial_{\mu} \phi  - V(\phi) \nonumber \\
-  \frac{B_{F}(\phi)}{4}F_{\mu\nu}F^{\mu\nu} + \sum_{j=n,p,e}
[\bar\psi_j iD\!\!\!\!/ \psi_j -
B_j(\phi)m_j\bar\psi_j\psi_j]\Biggr\}. \label{lagrangian}
\end{eqnarray}
Since we are going to consider couplings of $\phi$ that are essentially much stronger than 
gravitational, the stability of the model will require that $V(\phi)$ and the $B_i(\phi)$ functions 
have a minimum with respect to $\phi$. 
In what follows, we shall adopt the following 
ansatz,
\be
V(\phi)=\Lambda_0 + \fr{1}{2}\Lambda_2(\phi-\phi_0)^2+...;~~~~~~ 
B_i(\phi) = 1 + \fr{1}{2}\xi_i(\phi-\phi_i)^2+...,
\label{quadratic}
\ee
where ellipses stand for cubic, quartic etc. contributions around the minima. 
Here $\xi_i$, $\phi_0$ and $\phi_i$ are arbitrary dimensionless numbers;
$\Lambda_0$ and $\Lambda_2$ have dimensions of [Energy]$^4$ and we are tempted to choose 
$\Lambda_0$ to be equal to the current dark energy density to ``solve" the dark energy problem. 

A further simplification of the quadratic ansatz comes from the assumption that the proton and neutron 
$B_{p(n)}$ functions are mostly induced by the gluon $B$-function, and thus are approximately 
equal. With these simplifying assumptions, we can take
\be 
\phi_n\simeq \phi_p \equiv \phi_m; ~~~~~ \xi_n\simeq \xi_p=1; ~~~~~ \phi_0=0.  \label{assume}\ee
The normalization of $\xi_{p(n)}$ to one can be attained by rescaling $M_*$. In principle, a negative  value for $\xi$ is 
also possible, but in this section we shall restrict our discussion to positive $\xi$'s. 
Of course, the relations 
(\ref{assume}) are only approximate, and possible violations at the $\sim 1-10$ per mill level are 
naturally expected 
due to the nonzero quark and electromagnetic content of nucleons. The choice of $\phi_0=0$ can always be achieved 
by a constant shift of $\phi$. The ansatz (\ref{quadratic}) and (\ref{assume}) is very similar to the 
Damour-Polyakov model \cite{DP} (see also \cite{DN}), where all couplings to matter fields exhibit the same 
minimum. In the same vein, we assume the same minimum $\phi_m$ for 
$B_F(\phi)$ function. There are two important differences in our approach compared to the 
Damour-Polyakov models: we take $M_*$ to be much smaller than the 
Planck mass, and introduce a self-interaction potential that has a {\em different} 
minimum than the minimum of the $B_i(\phi)$ functions.

In this section we disregard higher-order nonlinear corrections to $ V_{\rm eff}$, postponing their 
discussion to Section 4. Furthermore, we assume a region of 
relatively uniform matter density $\rho$. 
In such regions, the scalar field equation of motion
takes the following form 
\be 
M_*^2\Box \phi +
\frac{\partial V_{\rm eff}}{\partial \phi}  =0,
\label{fieldeq} 
\ee 
where the effective potential is given by 
\be
V_{\rm eff} = \Lambda_0 + \fr{1}{2}\Lambda_2\phi^2 + \fr{1}{2}(\phi-\phi_m)^2\rho.
\label{ansatz}
\ee
 This potential creates the minimum for the scalar field at 
\be
\phi_{\rm min} =  \phi_m\frac{\rho}{\rho+\Lambda_2},
\label{phimin}
\ee
and the physical (canonically normalized) excitation $\varphi$ around this minimum has a mass
\be
m_{\rm eff}^2(\rho)  = \fr{\Lambda_2}{M_*^2} + \fr{\rho}{M_*^2} \equiv  
\fr{1}{\lambda_{\rm eff}^2}.
\label{meff}
\ee
By definition, the longest range for the $\varphi$-mediated force is achieved in vacuum at $\rho=0$.  
It is instructive to present a numerical formula for $\lambda_{\rm eff}$ at $\rho \gg \Lambda_2$:
\be
\lambda_{\rm eff} = 7\times 10^{-3}~{\rm cm}\times \fr{M_*}{\rm 1~TeV}~
\left(\fr{10^{24}~{\rm GeV~cm}^{-3}}{\rho}\right)^{1/2},
\label{range}
\ee
which shows that for an extreme case with a weak-scale $M_*$ and terrestrial matter densities
the range of the force falls under one millimeter.  

If  the spatial extent of the mass distribution is much larger than 
the Compton wavelength of the physical excitations of $\phi$
the effective interaction with a ``test" nucleon takes the following form,
\be
\label{Lint}
{\cal L}_{\rm int} = - m_N\bar NN\left( 1 + \frac{\phi_m^2\Lambda_2^2}{2(\Lambda_2+\rho)^2}-
 \frac{\varphi}{M_*}\fr{\phi_m\Lambda_2}{(\Lambda_2+\rho)} + \frac{\varphi^2}{2M_*^2}   \right),
 \ee
from where we can read a $\rho$-dependent mass of a nucleon, 
\be
m_{N \rm eff} = m_N\left( 1 + \frac{\phi_m^2\Lambda_2^2}{2(\Lambda_2+\rho)^2}\right),
\label{mneff}
\ee
and the scalar-field-corrected Newtonian interaction potential between two nucleons separated by distance $r$,
\be
U(r) = G_N \fr{m_N^2}{r}\left(1 + 
\exp(-m_{\rm eff}r)\times\frac{2M_{\rm Pl}^2}{M_*^2}\frac{\phi_m^2\Lambda_2^2}{(\Lambda_2+\rho)^2}\right).
\label{1/r}
\ee

Perhaps the most interesting case to consider is $\Lambda_2\gg \rho$ for low density environments,
such as {\em e.g.} the interstellar medium, and 
$\Lambda_2\ll \rho$ for high density environments such as stars and planets.
In that case, the change in the nucleon mass and the fine structure constant  
can be expressed as 
\begin{eqnarray}
\nonumber
\frac{\Delta m_N}{m_N} &=& \frac{m_{Nr}-m_{Nd}}{m_N}\simeq \frac{\phi_m^2}{2};\\
\fr{\Delta \alpha}{\alpha} &=& \frac{\alpha_r -\alpha_d}{\alpha} = -\frac{\xi_F \phi_m^2}{2},
\label{change}
\end{eqnarray}
and we assume that $\phi_m^2$ and $\xi_F\phi_m^2$ are much less than one. 
Notice that $\xi_F$ can be as large as  $\xi_F\sim O(100)$ without 
violating the assumption that $\xi_n\simeq\xi_p\simeq 1$.

\section{Experimental constraints on the model}
\setcounter{equation}{0}

All experimental constraints on the model described by 
(\ref{Lint}) can be divided into two broad categories.
The constraints coming directly from the quadratic couplings of $\phi$ to matter to 
a large extent do not depend on the position of the minimum of $\phi$ 
and on whether this minimum can be reached for a realistic size of an overdensity in question. 
The second group of constraints follows from the linear coupling of $\phi$ to matter, which are 
very sensitive to the position of $\phi$ and on the size of the overdensity. 

{\em General remarks on chameleon-type models.}
Before we proceed with the analysis of our model, we would like to make several remarks regarding 
chameleon models. It has been shown that an appropriate
choice of self-interaction potentials \cite{cham1} relaxes gravitational and astrophysical 
constraints on the density dependent interactions of the scalar field \cite{ekow} 
caused by a shift in the field $\phi$. 
Density dependent couplings could in principle cause a shift in the field value 
as well as its mass.  However, the magnitude of the shift will depend sensitively on
the local density and the length scale over which the shift occurs.
For the chameleon mechanism to work, we must require that $\partial V_{\rm eff}/\partial \phi
\gg M_*^2 R^{-2} \delta \phi$ or $m^2 R^2 \gg 1$, 
where $R$ is the characteristic scale of the density enhancement
and $\delta \phi$ is shift in $\phi$ from the low density (cosmological) solution of $\phi$ 
to the local value. When this condition is satisfied, the field and its mass inside 
the overdensity and away from the boundary will
be determined by the solution of a spacially homogeneous equation $\partial V_{\rm eff}/\partial \phi = 0$. 

To begin with, let us consider models with a linear coupling to density
\be
V_{\rm eff}  = V(\varphi) + \beta \rho \varphi, 
\ee
where we use the canonical normalization for the kinetic term of $\varphi$ and choose $\beta = M_*^{-1}$.
In chameleon models with a quartic potential, $V(\varphi) = \lambda \varphi^4$, 
the shift in $\varphi$ is tiny as long as $\lambda$ is not tremendously small. 
The effective mass when $\partial V_{\rm eff}/\partial \varphi = 0$
is $\lambda^{1/6} (\beta\rho)^{1/3}$ and for $\lambda\sim O(1)$ 
is far greater than $ R_\oplus^{-2}$.
In this case, as for all stiff potentials, the local environment
determines the field dependent couplings.  
Another example is the quintessence-like potential $V(\varphi) = M^5/\varphi$ with $M \sim 10^{-24} M_{\rm Pl}$.
The cosmological background solution for $\beta \sim M_{\rm Pl}^{-1}$ 
gives $\varphi \sim M_{\rm Pl}$ so that the potential and 
cosmological density are of the same order of magnitude.
Naively, the local solution of $\partial V_{\rm eff}/\partial \varphi = 0$ would yield 
$\varphi^2 = M^5 M_{\rm Pl} / \rho_\oplus \sim 10^{-27} M_{\rm Pl}^2$ taking $\rho_\oplus \sim 10 ^{-93} M_{\rm Pl}^4$.
However in this case, the gradient term in eq. (\ref{fieldeq}) dominates and the correct solution
for the local value of $\phi$ is a small shift from the background value of order
$\delta \varphi \sim \rho_\oplus R_\oplus^2/M_{\rm Pl}$ or 
$\delta \varphi \sim gR_\oplus M_{\rm Pl} \sim 10^{-9} M_{\rm Pl}$ where
 $g$ is the local acceleration on the 
Earth's surface.
In this case, the chameleon mechanism is not operative.
It can be restored if one adds a constant, $M^4$ to $V(\varphi)$. By doing so, the mass scale, 
$M$, can be made significantly  smaller 
$M \sim 10^{-30} M_{\rm Pl}$ \cite{cham2} and 
now the background solution for $\varphi$ yields a very small value $\varphi \sim 10^{-15} M_{\rm Pl}$.
In this case, the gradient terms can be safely neglected and the local solution for $\varphi$ is indeed given
$\partial V_{\rm eff}/\partial \varphi = 0$.
A similar argument can be made to show that the local density has virtually no effect on the 
background field value for potentials of the form $V(\varphi) = -\mu^4 \ln (\varphi/M_{\rm Pl})$
with $\mu \sim 10^{-3}$ eV that was discussed recently in \cite{nk}.

{\em Exact solutions for spherical (under)overdensities.}
The model based on the effective potential (\ref{ansatz}) considered in this paper is simple to analyze, as the field equations are linear. 
This allows us find an analytic solution for a spherical region of constant density,
and consider both large and small $M_*$ limits.  
The general form of the solution for $\phi$ as the function 
of radius for a spherical region of density $\rho_1$ of maximal extent $R$ surrounded by the infinite 
region of density $\rho_2$ takes the following form: 
\be
\label{solution}
\phi(r)=\left\{\begin{array}{c} \phi_1+\fr{A}{r}\sinh(m_1r),~~~~r<R;\\\\
                                \phi_2+\fr{B}{r}\exp(-m_2r),~~~~r>R,
                                \end{array}\right.
\label{rho12}
\ee
where the constants of integration
\begin{eqnarray}
\label{AB}
~~~~~~~~~~A = \fr{(\phi_2-\phi_1)(1+Rm_2)}{m_2\sinh(m_1R)+m_1\cosh(m_1R)};\\
B =\exp(m_2R)\times\fr{(\phi_2-\phi_1)(\sinh(m_1R)-Rm_1\cosh(m_1R))}{m_2\sinh(m_1R)+m_1\cosh(m_1R)};
\nonumber
\end{eqnarray}
are uniquely determined by boundary conditions, for which the physical choice is $\phi(\infty)=\phi_2$ and 
$\phi'(0)=0$.  
In these formulae, $m_1$, $m_2$, and $\phi_1$, $\phi_2$ are the mass and vacuum expectation values
calculated according to (\ref{meff}) and (\ref{phimin}) for $\rho = \rho_1$, $\rho_2$. Eq. (\ref{solution}) 
is a  generalization of the solution previously found in Ref. \cite{ekow}.

Adapting this solution to the case of $\rho_1=\rho_\oplus$ and $R=R_\oplus$, it 
is easy to see that in the limit of large $M_*$, or more precisely 
$m_1 R_\oplus\simeq \rho_\oplus^{1/2}M_*^{-1}R_\oplus\ll 1$, the solution simplifies to a
quadratically rising function on the inside and a $1/r$-falling function on the outside
(for simplicity, we also take $\Lambda_2,\rho_2 \to 0$ leading to $\phi_2 = 0$),
\be
\phi(r)=\left\{\begin{array}{c} -\phi_1 \left(\fr{m_1^2r^2}{6}-\fr{m_1^2R^2}{2}\right),~~~~r<R;\\\\
                               \phi_1 \fr{m_1^2R^3}{3r},~~~~r>R.
                                \end{array}\right.
\label{simplified}
\ee
It is easy to see that this solution exactly follows the gravitational potential profile,
with $\phi_1=\phi_m$. 
The change in the nucleon mass between $r=R_\oplus$ and spatial infinity 
is 
\be
\fr{\Delta m_N}{m_N} = \fr12(\phi^2(\infty)- \phi^2(R_\oplus)) \sim (g R_\oplus) \times \fr{ M_{\rm Pl}^2\phi_m^2}{M_*^2} 
\sim (10^{-9}-10^{-8})\times \fr{\phi_m^2 M_{\rm Pl}^2}{M_*^2}. 
\ee
which is exactly what one anticipates in a linearized scalar-tensor theory of gravity
with the relative strength of spin-0 to spin-2 exchange given by $2M_{\rm Pl}^2\phi_m^2/M_*^2$.
Furthermore, this limit, namely $\rho_\oplus^{1/2}M_*^{-1}R_\oplus\ll 1$ is
guaranteed in our model as long as $M_* > 10^{13}$ GeV. 
This regime is of no further interest to us in this paper, 
as it has been investigated in a number of previous publications. 

The opposite regime is achieved 
when $m_1R\gg 1$, in which case the interior solution quickly adjusts to $\phi=\phi_1$ 
for $r<R$. Moreover, if in addition $m_2R\gg 1$, the
spherical symmetry of the problem becomes irrelevant, and the solution degenerates 
into $\phi$ being frozen to its respective minima, $\phi_1$ and $\phi_2$, 
everywhere in space except for a 
small region near the surface separating two density regions. Defining this surface 
as $z=0$, we can write down a simplified form of (\ref{rho12}) which now takes 
the form:
\be
\phi(x)= \left\{\begin{array}{c} \phi_1+\fr{(\phi_2-\phi_1)m_2}{m_1+m_2}\exp(m_1z),~~~~z<0;\\\\
                                \phi_2+\fr{(\phi_1-\phi_2)m_1}{m_1+m_2}\exp(-m_2z),~~~~z>0.
                                \end{array}\right.
\ee
A smooth boundary would lead to an adiabatic adjustment of $\phi$ between its minima
provided that the Compton length of the scalar is 
much less than the characteristic scale of density change. For the atmosphere 
this scale is given by $\Delta_{atm}
= 1/|d\log(\rho_{atm}/\rho_\oplus)/dz|\sim 1$km, and the adjustment of the scalar field will occur as long as
\be
\lambda_{{\rm eff}}(\phi_{atm}) \ll\Delta_{atm},
\label{atm}
\ee
ensuring that in a terrestrial laboratory environment $\phi$ 
is exponentially close to $\phi_{\rm min}$
(\ref{phimin}). Neglecting the small $\Lambda_2$-proportional contribution to the scalar field mass, we
find that the condition (\ref{atm}) implies that
\be
M_*\ll 1{\rm km}\times \rho_{atm}^{1/2}~~ \Longrightarrow ~~ M_*\ll 10^{9}~{\rm GeV}
\label{choice}
\ee
Clearly, this choice of $M_*$ selects scalar models that are significantly more 
strongly coupled than $M_{\rm Pl}$-normalized models and therefore
the behavior of $\phi$ on Earth will depart drastically from the gravitational potential 
$\sim g_{00}-1$. Next we explore whether the choice of strong coupling (\ref{choice})
can survive gravitational and astrophysical constraints.


{\em Astrophysical constraints.} First we discuss the astrophysical constraints on the model which 
employs a $\phi^2$ coupling to photons and nucleons (the linear coupling is suppressed 
by $(\Lambda_2/\rho_1)^2$ and is assumed to be $\ll 1$). It is clear that 
the quadratic coupling will be less severely constrained than a linear coupling by the thermal 
emission rate of $\phi$-quanta from the hot interiors of stars. Indeed, the overall emission rate scales as $M_*^{-4}$ rather than 
$f_a^{-2}$ as one would routinely find in an axion-type model. As a result, instead of 
a lower limit to $f_a$ or order $10^{9}-10^{10}$
GeV, we expect to find a much more relaxed bound on $M_*$, 
of the order of the electroweak scale. 

Let us calculate the emissivity of $\phi$ quanta due to pair annihilation of photons. 
The amplitude for this process is induced by the $\xi_F\phi^2F_{\mu\nu}F^{\mu\nu}$ term in the Lagrangian,
leading to a cross section for this process in the center of mass frame,
\be
\sigma_{\gamma\gamma\to \phi\phi} = \fr{\xi_F^2}{M_*^4}\,\fr{\omega^2}{32\pi}.
\label{sigma}
\ee
This cross section results in an energy loss (Energy/volume/time) for a thermalized gas of photons at the level of 
\be
\Gamma_{\gamma\gamma\to \phi\phi} = n_\gamma^2\langle 2 \omega \sigma_{\gamma\gamma\to \phi\phi}\rangle 
=\frac{\zeta(3)\pi}{63}\,\fr{\xi_F^2T^9}{M_*^4}\simeq 0.06\times \fr{\xi_F^2T^9}{M_*^4}.
\label{rate}
\ee
Comparing this to the typical limit on $\Gamma=\epsilon_x\rho_{core}<10^{-14}$MeV$^5$ that 
follows from the constraints on the emissivity of light particles in cores of supernovae \cite{Raffelt},
\be
\epsilon_x \la 10^{19}~{\rm erg~g^{-1}~s^{-1}~~at~~}\rho_{core}=3
\times 10^{14}~{\rm g~cm^{-3}},~~T_{core}=30\,{\rm MeV},
\label{SNparameters}
\ee
we obtain a typical sensitivity to the coupling of $\phi$ to photons, 
\be
M_*\xi_F^{-1/2}~ \ga~ 3~{\rm TeV}.
\label{astro_limit}
\ee
This limit is admittedly not very precise, as it is quite sensitive to the 
temperature of the core, and more conservative assumptions about $T_{core}$  may result in relaxation 
of (\ref{astro_limit}) by a factor of a few. 
Other channels of $\phi$-production from light species, such as $\gamma e \to e \phi \phi$ or 
$e^+e^- \to \phi\phi$ will be 
further suppressed by the smallness of electromagnetic couplings or by the ratio of $m_e/T$. 

The constraint (\ref{astro_limit}) is not far the limits on the 6-dimensional Planck scale $M_6$ in
models with two large extra dimensions where gravity is allowed to 
propagate \cite{ADD,ADD_SN,ADD_SN1} in extra dimensions. 
This is not a 
total coincidence: in models with two large extra dimensions the total emissivity of Kaluza-Klein gravitons
also scales as $M_6^{-4}$. We note that due to the high power of 
temperature in the emission rate (\ref{rate}) the supernovae constraints are expected to be 
superior to other astrophysical constraints from energy loss mechanisms derived from the considerations 
of red giants, old neutron stars, etc. (See Ref. \cite{Raffelt} for further details).  

Similar considerations can be applied to the bremsstrahlung-like emission process 
$N+N\to N+N+\phi+\phi$, where again pairs of 
$\phi$ are emitted. This process is important because the density of neutrons inside 
the core is rather large. Here, instead of performing a detailed calculation 
which is perhaps not warranted, we use a simple Weizsacker-Williams-type estimate for the energy loss.
Specifically, we estimate the probability of energy loss in a collision of two nucleons to be
the product of elastic nucleon-nucleon cross section $\sigma_{NN}$ and the probability of an emission of 
certain amount of energy into $\phi$-quanta by an initial or final state nucleon,
\be
\langle E \sigma v \rangle \sim \fr{1}{12 \pi^4} \fr{T^3m_N^2}{M_*^4} \left(\fr{T}{m_N}\right)^{1/2} \sigma_{NN},
\ee
which leads to an energy loss of 
\be
\Gamma_{NN\to NN \phi\phi} \sim \sigma_{NN}\times \fr{n_N^2T^{7/2}m_N^{3/2}}{12 \pi^4 M_*^4},
\label{Gbrem}
\ee
where and $n_N$ is the number 
density of neutrons $\simeq \rho/m_N$. For the relevant range of energies $\sigma_{NN}$ can be taken on the order of 
25 mbn \cite{ADD_SN1}. Using the same parameters as  before (\ref{SNparameters}), we arrive at the constraint on
$M_*$,
\be
M_* \ga 15 ~ {\rm TeV},
\label{astro_limit2}
\ee
which is very similar to (\ref{astro_limit}). 
With these constraints, we conclude that the 
effective range of $\varphi$-force in terrestrial environment (\ref{range}) 
can indeed be as short as a millimeter. 

{\em Gravitational force constraints.} Gravitational force constraints on the model are by far the 
most complicated as they depend very sensitively on the effective distance range this 
force is probed at.  

Could the laboratory measurements of the gravitational 
force improve over the astrophysical bounds (\ref{astro_limit})? First, we look at the exchange by 
two quanta of $\varphi$ that does not depend on $\phi_{\rm min}$. Such an exchange leads to a
$1/r^3$ potential, 
\be
V= -\fr{1}{r^3}\, \fr{m_N^2}{64 \pi^3 M_*^4},
\label{rcube}
\ee
which is limited by recent searches for deviations from the 
gravitational $1/r$ behavior 
at short distances \cite{Wash1}. 
Specifying the constraints on phenomenological coefficient $\beta_3$ from Ref. \cite{Wash2} to 
our model prediction (\ref{rcube}), we arrive at
\be
\label{grav_quadr}
V = -\beta_3 \fr{G_N m_N^2}{r} ~\fr{\rm 1~mm^2}{r^2}~~{\rm with}~~\beta_3<
 1.3\times 10^{-4}~~ \Longrightarrow ~~ M_*> 2~{\rm TeV},
\ee
which is very close to the astrophysical bounds (\ref{astro_limit}) and (\ref{astro_limit2}).
We note again that the transition from a $1/r$ to a $1/r^3$ potential that may occur in our model 
at short distances is very similar to the transition expected in theories with two large 
extra dimensions. 

Constraints from the Yukawa part of (\ref{1/r}) are somewhat less straightforward to implement. 
For a range of  $O(10^{-2}-1~$cm$)$ in a medium, 
the constraint on its strength \cite{Wash2} specialized to our case with the use of (\ref{1/r}) takes the 
following form 
\be
\fr{\phi_m^2m_N^2\Lambda_2^2}{4\pi M_*^2\rho^2} \la {\rm few} \times 10^{-40},
\ee
or
\be
\fr{\phi_m^2}{10^{-6}}\times \left(\fr{\rm TeV}{M_*} \right)^2\times\left(\fr{\Lambda_2}{\rm eV^4}\right)^2
\la 10^{12}-10^{13},
\label{grav_lin1}
\ee
where we took $\rho \simeq 10$ g/cm$^3$ for the density of molybdenum used in experiments of Ref. \cite{Wash1}.
Perhaps an even more convenient from of the 
same constraint arises when we trade $\Lambda_2^{1/2}/M_*$ for $1/\lambda_{vac}$, the range of 
$\phi$-force in the vacuum,
\be
\fr{\phi_m^2}{10^{-6}}\times \left(\fr{M_*}{\rm TeV} \right)^2\times\left(\fr{\rm km}{\lambda_{vac}}\right)^4
\la 10^3-10^4.
\label{grav_lin2}
\ee
Constraints (\ref{grav_lin1}) and (\ref{grav_lin2}) do not look intimidatingly stringent, 
and indeed can be satisfied by
an appropriate choice of $M_*$, $\phi_m$ and $\Lambda_2$. It is also important to note that these 
constraints can be satisfied with a relatively short-range $\phi$-mediated force in vacuum, that could 
be shorter in range than {\em e.g.} typical distances within the solar system.

{\em Clock comparison constraints and constraints on the variations of 
couplings} are of particular interest to us in this paper. The now classic comparison of atomic clocks at 
an altitude of $R_{orbit}= 10^4$km with clocks on the ground have 
produced the limit of $2\times 10^{-4}$ on possible deviations from predictions of general 
relativity \cite{Vessot} (as quoted in the review \cite{Will}),
\be
\fr{|\Delta \omega_{\rm H}|}{\omega_{\rm H}} \la 2\times 10^{-4}\times |\Phi(R_\oplus)-\Phi(R_{orbit})|
\sim 5\times 10^{-13},
\ee
where $\Delta \omega_{\rm H}$ is the extra frequency shift of the Hydrogen maser 
added to the shift predicted by general relativity, and $\Phi(r)$ is the gravitational potential at
distance $r$ from Earth's center. 
In our model, the difference between clock 
frequency on the ground and in orbit would receive 
an additional correction from the difference of coupling constants and masses caused by 
$\Delta \phi$: 
\be
\fr{\Delta \omega_{\rm H}}{\omega_{\rm H}}=\fr{\Delta(\alpha^4 m_e^2g_p m_p^{-1} )}{\alpha^4 m_e^2g_p m_p^{-1}}
=-\left(\fr{\Delta\phi^2}{2}\right)\times \left(1 -2 \xi_e + 4\xi_F\right),
\ee
where for simplicity we assumed the same scaling for $\Lambda_{QCD}$ and quark masses with $\phi$ 
which keeps the proton $g$-factor $g_p$ fixed as a function of distance. The density of a medium surrounding the satellite 
is certainly very low, and thus it is tempting to take $\Delta\phi^2=\phi_m^2$. 
Note however, that if the in-medium 
range of the force is much shorter than a typical scale of a satellite, $L_{sat}\sim 1-10$m, 
with its average density being $M_{sat}/L_{sat}^3\sim 100{\rm kg}/{\rm m}^3\sim 0.1 {\rm g/cm^3}$,
the scalar field inside a satellite will roll back to its in-medium value, resulting in an 
exponential suppression of $\Delta\phi$. Therefore, we expect the clock comparison constraint 
to be at the level of
\be
\phi_m^2\exp(-2L_{sat}/\lambda_{\rm eff}) \times \left|1 -2 \xi_e + 4\xi_F\right| \la 10^{-12},
\label{clock}
\ee
noting that the precise amount of exponential suppression would also depend on the position of clocks 
inside the satellite. 
Although very powerful for $M_*\ga 10^8$ GeV when the exponential factor is of order one, 
eq. (\ref{clock}) is not particularly 
constraining for an interesting range of 1TeV$<M_*<100$TeV, as the value of $\phi$-field 
inside the satellite would be very close to that on the surface of the Earth.

Straightforward constraints on $\phi_m^2$ can be deduced from the comparison of coupling 
constants measured in cosmological settings and in the laboratory. Interpreting the null results of 
\cite{Petitjean}, $|\Delta\alpha/\alpha|<10^{-5}$, 
in terms of parameters of our model, we obtain
the constraint
\be
|\xi_F| \phi_m^2 < 2 \times 10^{-5}.  
\label{alpha_quas}
\ee
However, should one accept the criticism expressed in Ref. \cite{anti-Petitjean}, and interpret
the result in \cite{Webb} as a non-zero value of $\Delta \alpha$, then one predicts 
$\xi_F \phi_m^2 \sim 10^{-5} $, and the sign of $\xi_F$ comes out to be positive
(for positive $\xi_F$, the low density environments have a larger coefficient in front of 
$F_{\mu\nu}^2$ and therefore a smaller value of $\alpha$ in agreement with \cite{Webb}). 
In a similar fashion, the indication of a non-zero $\Delta (m_e/m_p)$ \cite{Petitjean2} 
can be interpreted as a nonzero value for the $(\xi_e-\xi_p)\phi_m^2$ combination. 
Another interesting possibility occurs for a choice of parameters when 
$\rho_\oplus \gg \Lambda_2 \ga \rho_{qso}$, where $\rho_{qso}$ is the average density
in a quasar absorption system. In this case,  not only does one expect a variation in the couplings
measured in these systems, but one would expect a shift in $\alpha$ from one absorption 
system to another as function of their density\footnote{In this sense, the observed scatter in
$\alpha$ may be real and due to a local environmental effect. In this case, a variation of the
density within the absorber.  A related solution (which does not require new physics)
accounted for the scatter through another environmental effect, namely the variation of the 
isotopic abundance of Mg \cite{amo}.}.
 An additional analysis of  data in Refs. \cite{Webb}  and \cite{Petitjean} searching for
$\alpha(\rho)$ correlation might be warranted.

It is also very important to stress that the Oklo constraint on $\Delta \alpha$ does not 
carry any weight in our model. Indeed, the Oklo phenomenon obviously occurred in
large density environment, which means that $\phi=\phi_m$ with good 
accuracy back at the time the Oklo reactor was active, as well as it is now. 
As to the constraints from meteorites \cite{OPQ}, similarly to eq. (\ref{clock}), one expects an 
exponential suppression of the effect by the density of meteorite, as long its size is larger 
than the $\phi$-field penetration length, $L_{met} > \lambda_{eff}$.

{\em Cosmological constraints.} Cosmology can constrain the presence of new degrees of freedom in the 
Universe. Big Bang Nucleosynthesis can in principle impose a constraint on a number of new relativistic 
degrees of freedom that carries a comparable amount of entropy as photons or neutrinos. 
It is very easy to 
see, however, that even if $\phi$ is initially thermally excited, its decoupling occurs well before 
the neutrino decoupling because $ M_*^{-2}\ll G_F$. Since traditionally the BBN constraints are expressed 
in terms of the number of ``new neutrino species", 
we can immediately conclude that $\phi$ contributes to 
this number as $4/7$ or less and thus cannot be ruled out on the 
grounds of light element abundances \cite{OS}. 
It is interesting to note that the position of $\phi$ during the time of  BBN is close to $\phi_m$, as the 
energy density of nonrelativistic matter during BBN is comparable to the terrestrial $\rho$ making
the BBN sensitivity to $\Delta \alpha$ far less than (\ref{alpha_quas}). 

The late time evolution of $\phi$ can be rather uninteresting. The scalar field remains at $\phi=\phi_m$ 
until the moment when $\rho$ drops below $\Lambda_2$, after which it settles towards $\phi=0$. Since 
$M_*\ll M_{\rm Pl}$, the vacuum mass  
$m_0=\Lambda_2^{1/2}/M_*$ is much larger than the 
Hubble scale at which $\rho\sim \Lambda_2$ and any 
oscillations around the minimum are very efficiently damped. 
Thus, the transition from $\phi = \phi_m$ to to $\phi = 0$ was completed when $\rho \sim \Lambda_2 \sim {\rm  eV}^4$  at redshifts $z\sim 10^3-10^4$. 
The amount of energy released in 
such a transition is $\sim \phi_m^2 \Lambda_2$, and has a negligible 
effect on the expansion history, 
as $\phi_m^2$ is constrained to be much less than 1. 

An interesting possibility emerges when $\Lambda_2$ is comparable to the matter energy density 
at redshifts $z\sim 1$. Then, the evolution of the coupling constants in time 
towards the minimum at $\phi=0$ would 
occur on cosmological scales, while remaining frozen at 
$\phi\simeq \phi_m$ on galactic scales where the matter 
energy density is larger than the average cosmological energy density by many orders of magnitude. 
This choice of parameters serves as a concrete realization of 
scenario proposed in \cite{BarMot}, where it was suggested that the cosmological evolution 
of $\alpha$ may proceed in some sense independently from the evolution of $\alpha$ in higher 
density environments (see also \cite{Clifton} for a related discussion of $\Delta G_N$ as a function of cosmological 
environment). In our model, with $\Lambda_2 \sim \rho_c$, 
cosmological evolution of $\phi$ would not proceed indefinitely into the future but only up
to the moment when $\phi$ reaches its low-density minimum at zero. Once again, this
behavior is possible only because of the increased coupling strength to matter, $M_* \ll M_{\rm Pl}$.

\section{Nonlinear models}

In this section we re-introduce nonlinear corrections to $V_{eff}(\phi)$. 
In fact, several of such models were discussed in the 
``chameleon" literature \cite{cham1}-\cite{cham3}, 
where potentials $V_{eff} = V(\phi) + \phi\rho$ with different choices of 
$V(\phi)$ were extensively investigated. As we saw earlier, the quadratic coupling of 
$\phi$ to matter allows one to escape many stringent constraints, and we intend to keep 
this feature in this section. To make our discussion more concrete, we limit the form of the 
potential $V(\phi)$ to have only even powers of $\phi$ and require the coupling to matter 
to have the same extremum,
\be
\label{nonlin}
V_{eff} = \fr14\Lambda_4\phi^4+\fr12(p\Lambda_2+q\rho)\phi^2 + \Lambda_0.
\ee
Such a Lagrangian may result, for example, from a discrete symmetry $\phi\to -\phi$. 
To ensure overall stability, we must choose $\Lambda_4$ to be positive. 
$p$ and $q$ are taken to be $\pm 1$, and we maintain $\Lambda_2>0$. Another 
way to interpret the model (\ref{nonlin}) is to say that we allow for ``tachyonic"
value of the mass of $\phi$ in (\ref{meff}) but ensure an overall stability of the potential 
by introducing positive $\phi^4$ contribution. The cosmological consequences of late time phase transitions 
were explored in \cite{Pietroni}.

The most interesting model of this type arises from the choice of parameters that 
allow for the spontaneous breaking of the discrete symmetry, $p=-q$.
\be
p=-1,~q=+1 \quad\Rightarrow\quad  
\left\{
\begin{array}{c}\langle\phi^2\rangle =0
\quad {\rm for}\quad \rho>\rho_c= \Lambda_2,\\
\langle\phi^2\rangle = \Lambda_4^{-1}(\Lambda_2-\rho)
\quad {\rm for}\quad \rho<\rho_c= \Lambda_2.
\end{array}\right.
\ee
It is of course tempting to choose the critical value of matter density 
$\rho_c$ in excess of the average cosmological energy density, so that on average 
there is a breaking of the discrete symmetry in cosmological environments.
However, in matter overdensities of sufficient spatial extent the symmetry is restored, 
leading to the erasure of the cosmological vacuum expectation value $\langle \phi\rangle$. 
The difference between masses and couplings between dense and rarefied environments are given 
by 
\be
\frac{\alpha_r -\alpha_d}{\alpha}= -\xi_F \fr{\rho-\Lambda_2}{2\Lambda_4};
\quad \frac{m_r -m_d}{m}= \fr{\rho-\Lambda_2}{2\Lambda_4}
\label{alpha_nonlin}
\ee

In the broken phase (low-density environments), there is a linear coupling of the Higgs field $\phi - \langle \phi \rangle$ 
to matter, that creates a contribution to the Newtonian force between two test particles,
\begin{eqnarray}
U(r) &= &G_N \fr{m_N^2}{r}\left(1 +
\exp(-m_{\rm eff}r)\times\frac{2M_{\rm Pl}^2}{M_*^2}\frac{\Lambda_2-\rho}{\Lambda_4}\right)
\quad {\rm for}\quad \rho<\rho_c= \Lambda_2,;\\
m_{\rm eff}^2 &=&\fr{\Lambda_2-\rho}{M_*^2}.
\end{eqnarray}
Near the ``phase transition", $\rho\simeq \Lambda_2$, the range of the force mediated by $\varphi$ 
becomes infinite. 
In the unbroken phase, only the quadratic coupling of $\phi$ to matter survives, which 
as we saw before, significantly relaxes all constraints on the parameters. In other words, the
constraint (\ref{grav_quadr}) is still operative while (\ref{grav_lin1}) is no longer applicable,
provided that $\Lambda_2 < \rho_{\oplus}$. 


As a side remark, we note that there is another interesting 
spin-off of (\ref{nonlin}) with the following choice of parameters:
$p = -q = 1$ and $\rho_{star}\la \Lambda_2 \la \rho_{core}$, 
where $\rho_{core}$ is a typical density inside a 
stellar core during a supernovae explosion. In this set-up the scalar field keeps its zero expectation 
value everywhere except for the extremely dense environments where it is allowed to roll to
$\langle \phi^2\rangle \neq 0$. Therefore, for this choice of model parameters, there are
no consequences for the terrestrial and cosmological tests of $\Delta \alpha$. There is, however, 
an interesting possibility that the nucleon mass experiences a shift during the supernova 
explosion, which in turn may influence the energetics of the explosion and 
affect the total luminosity (see {\em e.g.} \cite{Fairbairn} where the
linear change in strong coupling is discussed in connection with supernova explosions). 
If the couplings of neutrons and protons to $\phi^2$ are different, 
this model can {\em enhance} the environmental dependence of the total SN type Ia luminosity, 
thus affecting the accuracy with which cosmological parameters can be extracted 
from supernovae data. The detailed discussion of such possibilities 
falls outside the scope of our paper. 

\section{Discussion}

{ \em On a possibility for new tests of $\alpha(\rho)$ and $m(\rho)$}

One possibility to search for the environmental change of masses and couplings caused by a 
change in density is to try and recreate a low-density environment in the laboratory. 
The best quality vacuums available today achieve a density at the level of $10^5$ particles/cm$^3$, which creates a
matter density comparable to eV$^4$. 
Taking $\rho_2\gg \Lambda_2\gg \rho_1$, and
using the generic solution (\ref{solution}), we calculate the resulting shift of $\phi$ and the 
change in the coupling constant between the center and the walls of a spherical chamber of radius $R$,
\begin{eqnarray}
\fr{\alpha(r=R)-\alpha(r=0)}{\alpha}~
\simeq ~\fr{\xi_F\phi_m^2}{2}\left\{\begin{array}{c}
\fr{1}{36}\left(\fr{R}{\lambda_{vac}}\right)^4~~~{\rm for}~~~ R/\lambda_{vac} \ll 1, \\\\
1~~~{\rm for}~~~ R/\lambda_{vac} \ga 1 .
\end{array}\right.
\label{Rvac}
\end{eqnarray}
Similar changes will be experienced by masses of particles. Notice that the parametric
dependence of (\ref{Rvac}) is very similar to (\ref{grav_lin2}), 
and plugging $\Lambda_2$ and $\phi_m$ that saturate this constraint we find that
\begin{eqnarray}
\fr{\alpha(r=R)-\alpha(r=0)}{\alpha}~
\sim ~\xi_F\left(\fr{{\rm TeV}}{M_*}\right)^2\times\left\{\begin{array}{c}
(10^{-17}-10^{-16})\times \left(\fr{R}{{1 \rm m}}\right)^4~~~{\rm for}~~~ R/\lambda_{vac} \ll 1, \\\\
(10^{-15}-10^{-14})\times\left(\fr{\lambda_{vac}}{1\rm m}\right)^4~~~{\rm for}~~~ R/\lambda_{vac} \ga 1 .
\end{array}\right.
\label{Rvac_num}
\end{eqnarray}
These shifts are extremely small, but perhaps are not so far away from the
modern capabilities of frequency measurements 
that can be sensitive to the relative shifts as low as $10^{-15}$ \cite{Hansch}. 
Further gain in sensitivity can be achieved with substances that have a large 
enhancement factors that connect $\Delta\omega/\omega$ with $\Delta \alpha /\alpha$ \cite{Flambaum}. 

To put (\ref{Rvac_num}) in perspective, we compare this with the result
obtained  for $\Delta\alpha/\alpha$ 
between two points separated vertically by $\sim 1$m in the original Bekenstein model with
a massless scalar field. 
In that model, the coupling 
of $\phi$ to matter is linear and very weak, and the equivalence principle tests require that
$\zeta_F^2 (M_{\rm Pl}/M_*)^2 < 10^{-6}$ (see e.g. Fig 2 of \cite{OP}). In the Bekenstein model,
$\Delta \alpha/ \alpha = \zeta_F \Delta \phi$, where $\Delta \phi$ is the difference in $\phi$
over 1m above the surface of the Earth. The solution for $\phi$ is given by
\be
\phi = {\zeta_m \rho r^2 \over 6 M_*^2} ,
\ee
where $\zeta_m$ is the coupling of the scalar to matter (this can be obtained easily
from the massless limit of the solution given in \cite{ekow}). In the Bekenstein model,
this is absent in the Lagrangian but is induced by the coupling of the scalar to
the electromagnetic field in the nucleon.  Roughly, one finds $\zeta_m \sim 10^{-4} \zeta_F$ \cite{OP}.
Therefore $\Delta \phi$ is simply 
\be
\Delta \phi = {\zeta_m \rho R_\oplus h \over 3 M_*^2} = {10^{-4} \zeta_F g h \over 3} 
\left(\fr{M_{\rm Pl}}{M_*}\right)^2
\ee
where $h = 1$m is the vertical separation of the two points.
Putting all factors together one arrives at
\be
\fr{\Delta\alpha}{\alpha}(1{\rm m}) \sim (g\times 1{\rm m})\times 10^{-4}\zeta_F^2
\left(\fr{M_{\rm Pl}}{M_*}\right)^2
\la 10^{-26}~~~~~{\rm Bekenstein~ model},
\label{Bek_num}
\ee
where  $g$ is again the free-fall acceleration at the Earth's surface. This is well below any
detection sensitivity for the foreseeable future. The difference between 
(\ref{Rvac_num}) and (\ref{Bek_num}) is an enormous factor of nine orders of magnitude 
that can be traced back to the fact that equivalence principle is checked  at macroscopic distances
far better than gravity at distances under 1 mm. Chameleon models do not allow one to make 
a simple universal estimate of an allowed shift because they depend in a crucial way on the form of the 
self-interaction potential. Widely discussed chameleon models 
would not allow for a large spatial variation, even if one departs from the unviersal coupling of 
chameleon field to matter. This is again related to the extremely tiny values of chameleon field
relative to the Planck mass, and $O(1/M_{\rm Pl})$ size of the chameleon coupling to matter.

Estimates for $\Delta(m_e/m_p)$ similar to (\ref{Rvac_num}) involve a different factor, $\xi_e-\zeta_p=\xi_e-1$. 
It turns out that this factor can much larger than unity by up to two orders of magnitude. The 
difference comes about due to relaxed astrophysical constraints on $\xi_e/M_*^2$. Indeed, 
the rate of $e^++e^-\to \phi\phi$ is additionally suppressed relative to (\ref{rate}) by a factor of 
$(m_e/T)^2$, thus allowing for a much larger value of $\xi_e$. 

If a nonlinear model of the scalar field such as that described by Eq. (\ref{nonlin}) is realized, 
an artificially created  underdensity may lead to the shift of 
couplings according to (\ref{alpha_nonlin}). For the 
shift to occur, one needs $\rho_{r<R} < \Lambda_2 < \rho_{r>R}$ and $R> M_*\Lambda_2^{-1/2}$, which would allow 
for the broken phase to be created within $r<R$ volume.

{\em Astrophysical checks.} The possibility of an environmental dependence of coupling 
constants calls for new tests of $\Delta\alpha(\rho)$ and $\Delta(m_e/m_p)(\rho)$. 
In Bekenstein-type models where 
couplings evolve in time, there is a clear gain 
from testing $\alpha$ at maximally available 
redshifts. On the contrary, in the models discussed here, there is no gain in the size of the effect  
at large redshift for tests of $\Delta\alpha(\rho)$, and 
therefore one could search for new test sites within our galaxy. 
Atomic and molecular absorption lines in the interstellar medium 
could be an example where extremely narrow lines can be detected. 
Another beneficial aspects of testing for variations within our galaxy is the 
possible access to heavier elements 
where relativistic effects are significantly enhanced. It remains to be seen 
whether $O(10^{-5})$ sensitivity for $\Delta \alpha/\alpha$ achieved in the QSO 
absorption spectra can be improved upon using galactic lines. 

\section{Conclusions}

We have demonstrated that a scalar coupling to matter can be much stronger than the
gravitational  coupling in 
Damour-Polyakov type models. The quadratic  nature of coupling to matter allows one to escape 
the most prohibitive astrophysical and gravitational constraints, as only  pair-production or 
pair-exchange of $\phi$-quanta are allowed. The environmental dependence of masses and coupling 
constant can come about from the shift in the expectation value of $\phi$ between 
dense and rarefied environments. We have shown that such shifts could be at a detectable 
level, and could indeed be probed with astrophysical tests within our galaxy and in
laboratory clock comparison experiments. 

{\bf Acknowledgments.} The authors would like to thank Nemanja Kaloper and 
Nelson Nunes for discussions on chameleon-type models. 
M.P. would like to acknowledge useful conversations with 
Dave DeMille, Sara Ellison, Justin Khouri and Adam Ritz.   
The work of K.A.O.\ was partially supported by DOE grant DE-FG02-94ER-40823. 
The work of M.P. was supported in part
by NSERC, Canada, and research at the Perimeter Institute
is supported in part by the Government of Canada through NSERC and by the Province
of Ontario through MEDT.

\end{document}